\documentclass{article}

\usepackage{PRIMEarxiv}

\usepackage[utf8]{inputenc} 
\usepackage[T1]{fontenc}    
\usepackage{hyperref}       
\usepackage{url}            
\usepackage{booktabs}       
\usepackage{amsfonts}       
\usepackage{amssymb}  
\usepackage{amsthm}
\usepackage{amsmath}
\usepackage{nicefrac}       
\usepackage{microtype}      
\usepackage{lipsum}
\usepackage{fancyhdr}       
\usepackage{graphicx}       
\graphicspath{{media/}}     

\title{Transcranial ultrasound simulation with uncertainty estimation}

\author{
  Antonio Stanziola\textsuperscript{1}, José A. Pineda-Pardo\textsuperscript{2},  Bradley Treeby\textsuperscript{1}\thanks{bradley.treeby@ucl.ac.uk}\\
  \vspace{-0.5em}\\
  \textsuperscript{1}University College London, Gower Street, London, WC1E 6BT, UK\\
  \textsuperscript{2}HM CINAC (Centro Integral de Neurociencias Abarca Campal)\\Fundacio n Hospitales de Madrid, HM Hospitales, Madrid, Spain\\
}

\begin{document}
\maketitle

\vspace{-2em}

\begin{abstract}
Transcranial ultrasound simulations are increasingly used to predict \emph{in situ} exposure parameters for ultrasound therapies in the brain. However, there can be considerable uncertainty in estimating the acoustic medium properties of the skull and brain from computed tomography (CT) images. Here, we show how the resulting uncertainty in the simulated acoustic field can be predicted in a computationally efficient way using linear uncertainty propagation. Results for a representative transcranial simulation using a focused bowl transducer at 500 kHz show good agreement with unbiased uncertainty estimates obtained using Monte Carlo.
\end{abstract}

\section{Introduction}

Transcranial ultrasound simulations are increasingly used to predict \emph{in situ} exposure parameters for ultrasound therapies in the brain, including for blood-brain barrier opening \cite{deffieux2010numerical}, high-intensity focused ultrasound ablation \cite{mcdannold2019elementwise}, and transcranial ultrasound stimulation (TUS) \cite{lee2016transcranial}. Simulations are particularly important for treatment planning in TUS, as online monitoring methods (such as thermometry and acoustic radiation force imaging) are not yet sufficiently sensitive for use with low-intensity ultrasound therapies \cite{li2022improving}. 

The current approach for transcranial ultrasound simulation is to map the geometry and acoustic properties of the skull and brain from computed-tomography (CT) images \cite{marquet2009non}. However, there can be considerable uncertainty in the mapping from Hounsfield units to acoustic parameters \cite{webb2018measurements,webb2020acoustic}. Errors in the sound speed of the skull in particular, can result in changes to the amplitude and shape of the acoustic field inside the brain  \cite{vaughan2001effects,robertson2017sensitivity,montanaro2021impact}. In principle, these variations could be quantified using a Monte Carlo approach, where many simulations are performed using different acoustic property mappings, and the results of these simulations used to quantify the uncertainty. However, Monte Carlo estimation can take a large number of samples (i.e., simulations) to converge, significantly increasing the computational cost of model-based treatment planning.

In general, quantifying uncertainty is a challenging task for which many methods have been developed to avoid expensive Monte Carlo simulations. These range from the application of the Koopman operator \cite{gerlach2020koopman}, to density estimation \cite{papamakarios2016fast}. Here, we demonstrate how computationally efficient uncertainty estimation can be achieved using linear uncertainty propagation with a differentiable wave simulator, j-Wave \cite{stanziola2022j}.

\section{Linear uncertainty propagation}

For a given scanner and image acquisition and reconstruction settings, the mapping from a CT image in Hounsfield units to an image of mass density in kg.m$^{-3}$ can be performed with the aid of a calibration image of a test object with known density. Figure 1(a) shows an example of such a calibration acquired using a CIRS electron density phantom (Model 062M) using the scanner details described in \cite{caballero2019zero}. The error bars show the standard deviation in the image values averaged over each tissue equivalent electron density plug. A linear curve is then typically used to map from density to sound speed (or equivalently, two different linear mappings are used to map to density and sound speed from Hounsfield units \cite{marquet2009non}). Figure 1(b) shows a scatter plot of the density and sound speed of the skull samples measured in \cite{webb2018measurements,webb2020acoustic}, along with a linear fit and the 95\% confidence interval for the linear model. There is considerable spread in the measurement data, which may be attributed to both measurement noise (e.g., in the time-of-flight picking used to estimate the sound speed) and inherent variations in the mapping for different skulls and different regions of the skull. Here, we only consider the uncertainty in the mapping from density to sound speed, but in principle, it is also possible to propagate the uncertainty in the mapping from Hounsfield units to density, and to account for uncertainty in other acoustic and thermal parameters.

\begin{figure}
\centering
\includegraphics[width=0.8\textwidth]{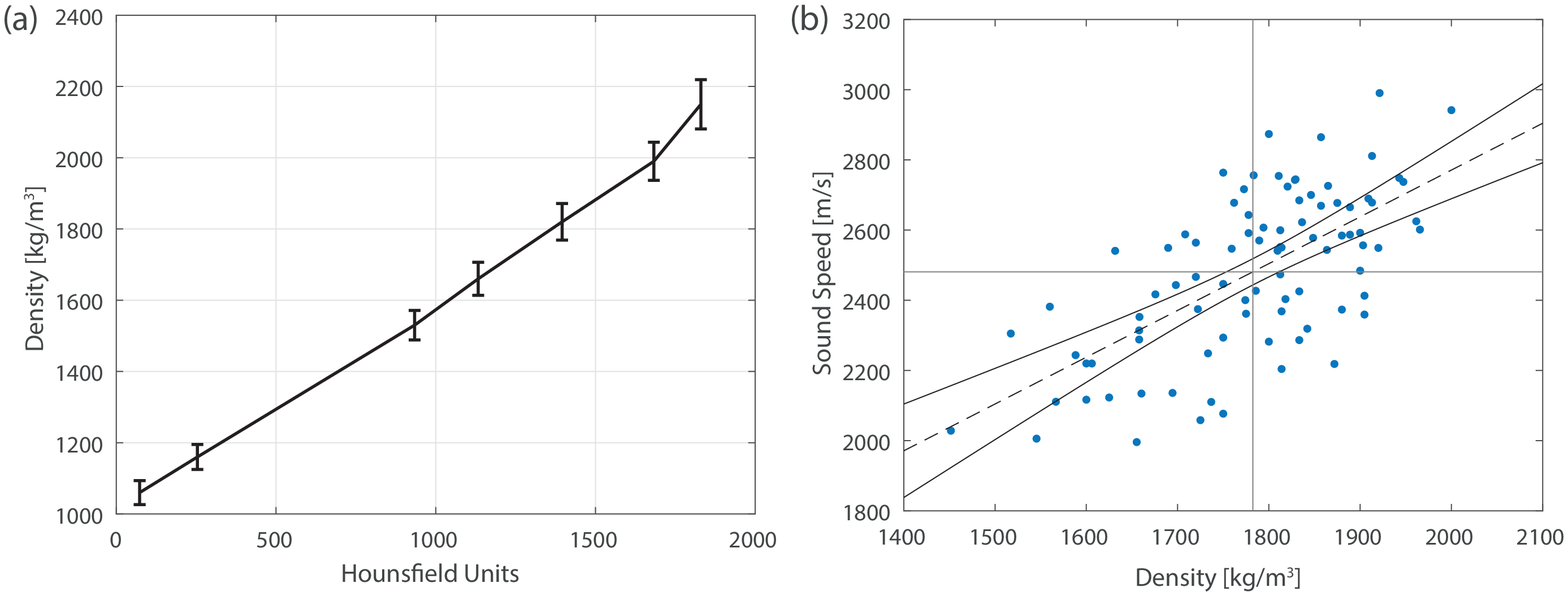}
\caption{\label{fig:uncertainty}(a) Conversion between Hounsfield units and mass density obtained using an image of a CIRS electron density phantom (Model 062M). The error bars show the standard deviation in the image values averaged over each tissue equivalent electron density plug. (b) Conversion between density and sound speed using the data from \cite{webb2018measurements,webb2020acoustic}. Each data point shows the density and sound speed measured for a specific skull sample (one outlier was removed from the data). The dashed line shows a linear fit to the data, and the solid lines show the 95\% confidence interval for the linear model. The cross hairs show the mean values.}
\end{figure}

Using the data shown in Fig. 1(b), it is assumed the mapping between density $\rho$ and sound speed $c$ is given by the linear relationship
\begin{equation}
    c(\alpha, \beta) = \hat{c} + \alpha (\rho - \hat{\rho}) + \beta,
\end{equation}
where $\hat c = 2481$ m.s$^{-1}$ and $\hat{\rho} = 1782$ kg.m$^{-3}$ are the mean values for the sound speed and the density in the data, respectively (shown by the cross-hair in Fig.\ 1(b)). Fitting the model on mean-zero data results in a linear fit with uncorrelated estimates for the $\alpha$ and $\beta$ parameters. While correlation of the input variables can be accounted for by linear uncertainty propagation, this simplifies the implementation. The least squares estimate of the slope and intercept are given by $\alpha_0 = 1.333$ and $\beta_0 = 0$ m.s$^{-1}$, with a standard deviation of $\sigma_\alpha = 0.168$ and $\sigma_\beta = 18.8$ m.s$^{-1}$.


Rather than performing a single acoustic simulation with a fixed mapping between density and sound speed (i.e., just using the values of $\alpha_0$ and $\beta_0$), to account for the uncertainty in the mapping, it is assumed that $\alpha$ and $\beta$ can be represented by uncorrelated Gaussian probability distributions. For a fixed problem (for example, a particular transducer and density map), the simulated acoustic pressure field $p$ can be written as a function of the parameter vector $\mathbf v = (\alpha, \beta)$
\begin{equation}
    p = f(\mathbf v) \enspace.
    \label{eq:function}
\end{equation}
Here, $f$ includes the wave simulation, and any other pre or post processing steps needed to generate the output pressure field. To propagate uncertainty, it is assumed that $\mathbf v$ is drawn from a normal distribution centered around $\mu_v$ with variance $\Sigma_v$, given by
\begin{equation}
    \mu_v = (\alpha_0, \beta_0), \qquad \Sigma_v = 
    \begin{pmatrix}
        \sigma_\alpha^2 & 0 \\
        0 & \sigma_\beta^2
    \end{pmatrix},
\end{equation}
that is, $\mathbf v = \mu_v + \varepsilon$, with $\varepsilon \sim \mathcal{N}(0,\Sigma_v)$.  If $\varepsilon$ is small enough, Eq.\ \eqref{eq:function} can be expanded as a Taylor series up to second order
\begin{equation}
    f (\mathbf v) \approx f(\mu_v) + J(\mathbf v - \mu_v), 
    \qquad
        \text{with } J = \left.\nabla f\right|_{\mathbf v = \mu_v} = \begin{pmatrix}
            \partial f/\partial \alpha\\ 
            \partial f/\partial \beta
        \end{pmatrix} \enspace,
\end{equation}
where $J$ is the Jacobian.

The truncated first-order Taylor series expansion of $f (\mathbf v)$ is a linear function of $\mathbf v$, thus inputs drawn from a normal distribution are mapped to outputs that are also normally distributed. The mean-value of the distribution is $ f(\mu_v)$ (in other words, the simulated pressure field using $\alpha_0$ and $\beta_0$), while the variance is given by \cite{giordano2016uncertainty}
\begin{equation}
    \mathrm{Var}(f(\mathbf v)) = J \Sigma_v J^T \enspace.
\end{equation}
Since the variables are uncorrelated, the formula for the variance simplifies to
\begin{equation}
    \mathrm{Var}(f(\mathbf v)) = \left(\left.\frac{\partial f}{\partial \alpha}\right|_{\mathbf v = \mu_v} \right)^2\sigma_\alpha^2 + \left(\left.\frac{\partial f}{\partial \beta}\right|_{\mathbf v = \mu_v} \right)^2\sigma_\beta^2.
    \label{eq:variance}
\end{equation}
If the Jacobian is non-singular, this estimate can provide a good approximation for the variance for small variations of the parameter vector $\mathbf v$. To estimate the normal distribution of the outputs, all that is needed is an efficient way to calculate $f(\mu_v)$ and the Jacobian vector of $f$ at $\mathbf v=\mu_v$.

\section{Numerical Experiments}

To demonstrate uncertainty estimation using Eq.\ \eqref{eq:variance}, an ultrasound simulation through a CT-derived acoustic property map was performed in the context of TUS. A CT image of the head was taken from the dataset described in \cite{caballero2019zero,miscouridou2022classical}, re-sampled to an isotropic resolution of 0.5 $\times$ 0.5 $\times$ 0.5 mm, and truncated to a 120 $\times$ 70 $\times$ 70 mm domain in the region of the primary motor cortex. The simulation uses speed of sound and density values derived from the CT image, while medium is considered lossless. The transducer was a focused bowl defined following \cite{aubry2022benchmark}, with a radius of curvature and aperture diameter of 64 mm, source pressure of 60 kPa, and driving frequency of 500 kHz. 

Simulations were performed using j-Wave \cite{stanziola2022j}, a differentiable wave solver, by running a time-domain simulation to steady state, and then recording the maximum pressure amplitude over several cycles. To avoid staircasing artefacts, the transducer was represented on the grid using the band limited interpolant \cite{wise2019representing}. The grid spacing gave 6 points per wavelength (PPW) in water, and the time step was set using a Courant-Friedrichs-Lewy number of 0.3.

To calculate the gradients $\partial f/\partial \alpha$ and $\partial f/\partial \beta$, forward-mode differentiation was used \cite{stanziola2021jaxdf,stanziola2022j}. In j-Wave, this can be performed in parallel with the unperturbed forward computation, at a cost of one forward computation for each partial derivative. Forward-mode differentiation, in contrast to the widely-used adjoint method, does not necessitate storing the intermediate computations. In the context of wave physics it has been used, for instance, to perform sensitivity-analysis of photonic devices \cite{hughes2019forward}. Alternatively, the gradients could also be approximated using a non-differentiable wave solver by replacing the gradients with a finite difference approximation. 

As a reference, the standard deviation of the simulation was also estimated via Monte Carlo, by running 200 simulations with samples of the parameter vector $\mathbf v$ (i.e., different values for the slope and intercept in the linear relationship between density and sound speed) drawn from its probability distribution.

\section{Results}

\begin{figure*}[t]
    \centering
    \makebox[\textwidth][c]{\hspace*{-0cm}
        \scalebox{.60}{\includegraphics{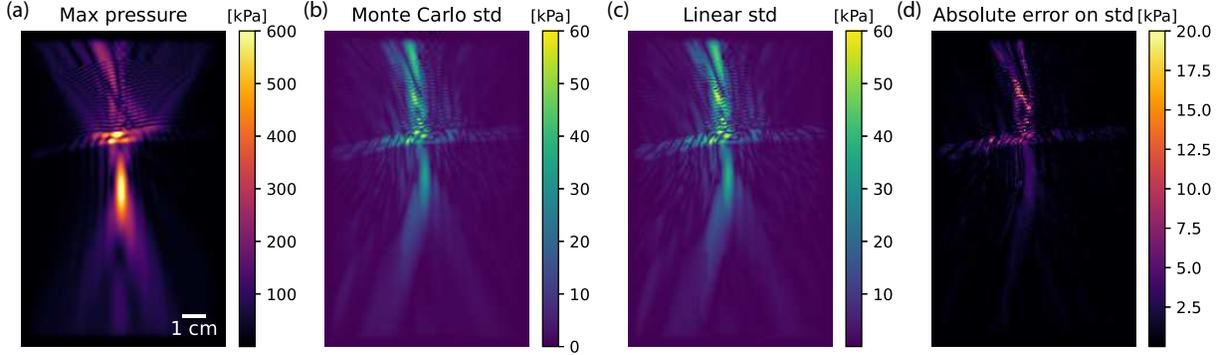}}
    }
 \caption{\label{fig:example_2d_maps} (a) The simulated pressure field, (b) the standard deviation (STD) of the simulation results using 200 Monte Carlo samples, (c) the standard deviation estimated by the linear uncertainty propagation method, and (d) the error between the estimated standard deviations.}
\end{figure*}

\begin{figure*}[t]
    \centering
    \makebox[\textwidth][c]{\hspace*{-0cm}
        \scalebox{.60}{\includegraphics{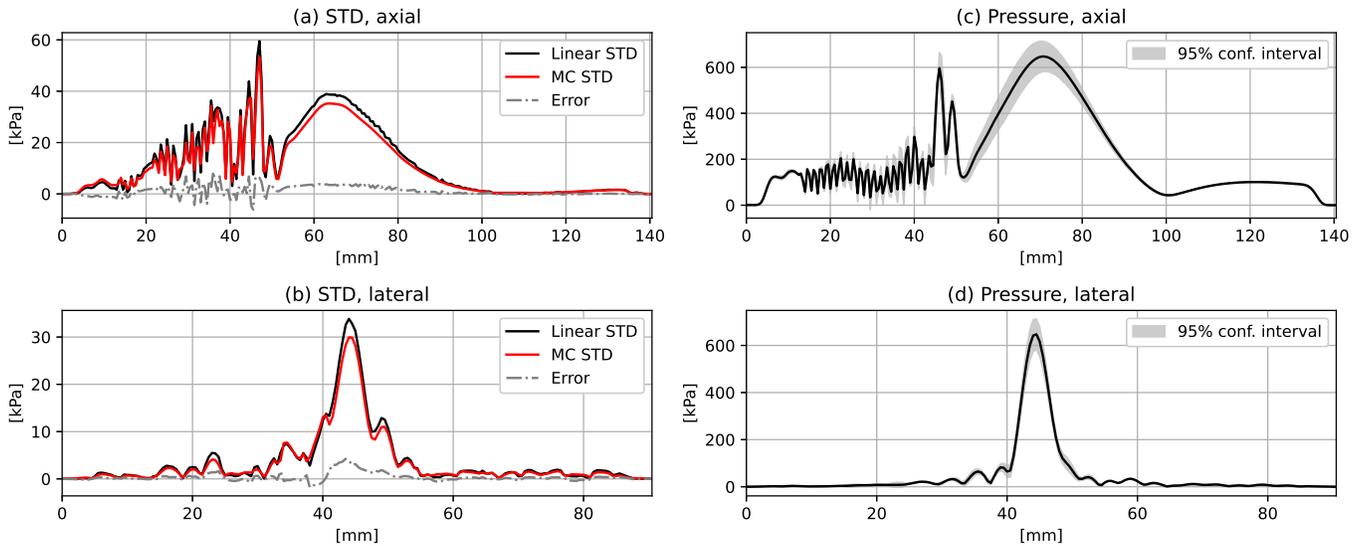}}
    }
 \caption{\label{fig:line_plots} The standard deviations (STD) estimated using Monte Carlo (MC) and linear uncertainty propagation (Linear), and the difference error between the two, for the (a) axial and (b) lateral lines passing trough the focus. Subplots (c) and (d) show the axial and lateral pressure profiles through the focus, along with the 95\% confidence interval (2-sigma) estimated using linear uncertainty propagation.}
\end{figure*}

The simulated acoustic pressure field for a plane passing through the focus is shown in Fig.\ \ref{fig:example_2d_maps}(a).
The standard deviation calculated using Monte Carlo is displayed in Fig.  \ref{fig:example_2d_maps}(b), while the standard deviation calculated using linear uncertainty propagation is displayed in Fig. \ref{fig:example_2d_maps}(c). The two uncertainty maps are in close agreement, both in terms of the structures shown and the size of the standard deviation estimate. The difference between the two, illustrated in Fig.\  \ref{fig:example_2d_maps}(d),  shows that the uncertainty is accurately estimated, particularly below the skull and close to the focal region, while the most noticeable discrepancies between the two are found inside the skull and in the reflected wave-field. The uncertainty estimates for lateral and axial profiles through the focus are shown in Fig. \ref{fig:line_plots}, highlighting how the algorithm predicts an error of up to 40 kPa near the focal region, which is very close to the Monte Carlo estimate.

\begin{figure*}[t]
    \centering
    \makebox[\textwidth][c]{\hspace*{-0cm}
        \scalebox{.60}{\includegraphics{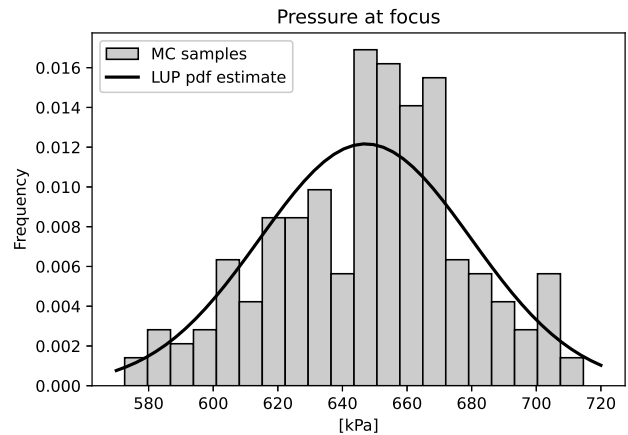}}
    }
 \caption{\label{fig:hist_at_focus} The histogram of the empirical distribution of the pressure value at the focus, estimated using 200 Monte Carlo runs, and the corresponding normal distribution estimated by linear uncertainty propagation (LUP).}
\end{figure*}

A representative example of the empirical probability distribution for the pressure value at the focus is shown in Fig. \ref{fig:hist_at_focus}. This shows a histogram of the pressure values at a single grid point for each of the 200 Monte Carlo simulations. The normal distribution predicted by the linear uncertainty propagation is shown on top. Despite the fact that the empirical distribution appears to be slightly skewed and therefore is deviating marginally from the supposed Gaussian distribution, the approximated distribution still satisfactorily explains the spread of possible pressure values for the given pixel.


\begin{figure*}[t]
    \centering
    \makebox[\textwidth][c]{\hspace*{-0cm}
        \scalebox{.70}{\includegraphics{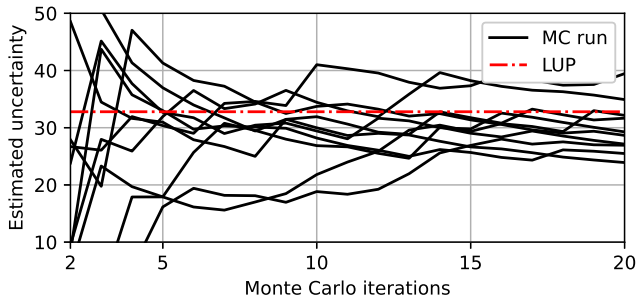}}
    }
 \caption{\label{fig:mc_convergence} Example of 10 different Monte Carlo (MC) runs and their convergence in estimating the standard deviation at the focus, along with the uncertainty estimated using linear uncertainty propagation (LUP).}
\end{figure*}

Figure \ref{fig:mc_convergence} shows the convergence of the estimated uncertainty using Monte Carlo for 10 different random seeds. With small numbers of iterations (simulations), the Monte Carlo estimate has a large amount of variability. While the Monte Carlo solution will eventually converge to an unbiased value, this can take, on average, a very large number of simulations. In contrast, linear uncertainty propagation only requires two simulations, so is computationally efficient. However, while for the simulation parameters used in this study it provides a good approximation of the true variance, for large values of $\varepsilon$, the biased nature of linear uncertainty propagation may start to become relevant.


\section{Summary}

We show that linear uncertainty propagation can be used to estimate uncertainty in simulated transcranial ultrasound fields when there is uncertainty in the medium property mapping between mass density and sound speed. This provides a computationally efficient way of putting error bars on acoustic simulations used for treatment planning due to uncertainty in the material properties. In future, we will explore accounting for additional uncertainties (for example, in the skull attenuation), the sensitivity of the predicted field to variations in different material properties, and evaluate the regime of validity of linear uncertainty propagation for different model parameters (e.g., transducer shape, frequency, target region, etc). 


Another potentially interesting extension would be to directly couple the acoustic simulation with a thermal simulation to estimate the uncertainty on the temperature rise induced by the acoustic field (this is relevant for thermal ablation, as well as safety for non-thermal therapies). As long as the thermal simulation is also written using a differentiable language, the algorithm outlined in this paper will still be applicable.

Lastly, note that the algorithm derived for estimating uncertainty is itself differentiable. This allows the predicted uncertainty map to be included in an other optimization loop and the definition of a cost function that depends on uncertainty that is still optimizable using gradient descent, as done for example in \cite{gerlach2020koopman} using a different method for estimating uncertainties.

\section*{Acknowledgments}

The authors would like to thank Taylor Webb and Kim Butts Pauly for provision of the raw data from \cite{webb2018measurements,webb2020acoustic} used to generate Fig.\ 1(b). This work was supported by the Engineering and Physical Sciences Research Council (EPSRC), UK, grant numbers EP/S026371/1 and EP/T022280/1.

\bibliographystyle{elsarticle-num}
\bibliography{references}

\begin{thebibliography}{10}
\expandafter\ifx\csname url\endcsname\relax
  \def\url#1{\texttt{#1}}\fi
\expandafter\ifx\csname urlprefix\endcsname\relax\def\urlprefix{URL }\fi
\expandafter\ifx\csname href\endcsname\relax
  \def\href#1#2{#2} \def\path#1{#1}\fi

\bibitem{deffieux2010numerical}
T.~Deffieux, E.~E. Konofagou, Numerical study of a simple transcranial focused
  ultrasound system applied to blood-brain barrier opening, IEEE transactions
  on ultrasonics, ferroelectrics, and frequency control 57~(12) (2010)
  2637--2653.

\bibitem{mcdannold2019elementwise}
N.~McDannold, P.~J. White, R.~Cosgrove, Elementwise approach for simulating
  transcranial mri-guided focused ultrasound thermal ablation, Physical review
  research 1~(3) (2019) 033205.

\bibitem{lee2016transcranial}
W.~Lee, H.-C. Kim, Y.~Jung, Y.~A. Chung, I.-U. Song, J.-H. Lee, S.-S. Yoo,
  Transcranial focused ultrasound stimulation of human primary visual cortex,
  Scientific reports 6~(1) (2016) 1--12.

\bibitem{li2022improving}
N.~Li, P.~Gaur, K.~Quah, K.~Butts~Pauly, {Improving in situ acoustic intensity
  estimates using MR acoustic radiation force imaging in combination with
  multifrequency MR elastography}, Magnetic Resonance in Medicine 88~(4) (2022)
  1673--1689.

\bibitem{marquet2009non}
F.~Marquet, M.~Pernot, J.-F. Aubry, G.~Montaldo, L.~Marsac, M.~Tanter, M.~Fink,
  {Non-invasive transcranial ultrasound therapy based on a 3D CT scan: Protocol
  validation and in vitro results}, Physics in Medicine \& Biology 54~(9)
  (2009) 2597.

\bibitem{webb2018measurements}
T.~D. Webb, S.~A. Leung, J.~Rosenberg, P.~Ghanouni, J.~J. Dahl, N.~J. Pelc,
  K.~B. Pauly, Measurements of the relationship between {CT} hounsfield units
  and acoustic velocity and how it changes with photon energy and
  reconstruction method, IEEE transactions on ultrasonics, ferroelectrics, and
  frequency control 65~(7) (2018) 1111--1124.

\bibitem{webb2020acoustic}
T.~D. Webb, S.~A. Leung, P.~Ghanouni, J.~J. Dahl, N.~J. Pelc, K.~B. Pauly,
  {Acoustic attenuation: Multifrequency measurement and relationship to CT and
  MR imaging}, IEEE transactions on ultrasonics, ferroelectrics, and frequency
  control 68~(5) (2020) 1532--1545.

\bibitem{vaughan2001effects}
T.~E. Vaughan, K.~Hynynen, Effects of parameter errors in the simulation of
  transcranial focused ultrasound, Physics in Medicine \& Biology 47~(1) (2001)
  37.

\bibitem{robertson2017sensitivity}
J.~Robertson, E.~Martin, B.~Cox, B.~E. Treeby, Sensitivity of simulated
  transcranial ultrasound fields to acoustic medium property maps, Physics in
  Medicine \& Biology 62~(7) (2017) 2559.

\bibitem{montanaro2021impact}
H.~Montanaro, C.~Pasquinelli, H.~J. Lee, H.~Kim, H.~R. Siebner, N.~Kuster,
  A.~Thielscher, E.~Neufeld, The impact of {CT} image parameters and skull
  heterogeneity modeling on the accuracy of transcranial focused ultrasound
  simulations, Journal of Neural Engineering 18~(4) (2021) 046041.

\bibitem{gerlach2020koopman}
A.~R. Gerlach, A.~Leonard, J.~Rogers, C.~Rackauckas, The {K}oopman expectation:
  An operator theoretic method for efficient analysis and optimization of
  uncertain hybrid dynamical systems, arXiv preprint arXiv:2008.08737 (2020).

\bibitem{papamakarios2016fast}
G.~Papamakarios, I.~Murray, Fast $\varepsilon $-free inference of simulation
  models with {B}ayesian conditional density estimation, arXiv preprint
  arXiv:1605.06376 (2016).

\bibitem{stanziola2022j}
A.~Stanziola, S.~R. Arridge, B.~T. Cox, B.~E. Treeby, {j-Wave: An open-source
  differentiable wave simulator}, arXiv preprint arXiv:2207.01499 (2022).

\bibitem{caballero2019zero}
J.~Caballero-Insaurriaga, R.~Rodr{\'\i}guez-Rojas,
  R.~Mart{\'\i}nez-Fern{\'a}ndez, M.~Del-Alamo, L.~D{\'\i}az-Jim{\'e}nez,
  M.~{\'A}vila, M.~Mart{\'\i}nez-Rodrigo, P.~Garc{\'\i}a-Polo, J.~A.
  Pineda-Pardo, {Zero TE MRI applications to transcranial MR-guided focused
  ultrasound: Patient screening and treatment efficiency estimation}, Journal
  of Magnetic Resonance Imaging 50~(5) (2019) 1583--1592.

\bibitem{giordano2016uncertainty}
M.~Giordano, Uncertainty propagation with functionally correlated quantities,
  arXiv preprint arXiv:1610.08716 (2016).

\bibitem{miscouridou2022classical}
M.~Miscouridou, J.~A. Pineda-Pardo, C.~J. Stagg, B.~E. Treeby, A.~Stanziola,
  {Classical and learned MR to pseudo-CT mappings for accurate transcranial
  ultrasound simulation}, IEEE Transactions on Ultrasonics, Ferroelectrics, and
  Frequency Control 69~(10) (2022) 2896--2905.

\bibitem{aubry2022benchmark}
J.-F. Aubry, O.~Bates, C.~Boehm, K.~{Butts Pauly}, D.~Christensen, C.~Cueto,
  P.~Gelat, L.~Guasch, J.~Jaros, Y.~Jing, R.~Jones, N.~Li, P.~Marty,
  H.~Montanaro, E.~Neufeld, S.~Pichardo, G.~Pinton, A.~Pulkkinen, A.~Stanziola,
  A.~Thielscher, B.~Treeby, E.~{van 't Wout}, {Benchmark problems for
  transcranial ultrasound simulation: Intercomparison of compressional wave
  models}, The Journal of the Acoustical Society of America 152~(2) (2022)
  1003--1019.

\bibitem{wise2019representing}
E.~S. Wise, B.~Cox, J.~Jaros, B.~E. Treeby, Representing arbitrary acoustic
  source and sensor distributions in fourier collocation methods, The Journal
  of the Acoustical Society of America 146~(1) (2019) 278--288.

\bibitem{stanziola2021jaxdf}
A.~Stanziola, S.~Arridge, B.~T. Cox, B.~E. Treeby, {A research framework for
  writing differentiable PDE discretizations in JAX}, Differentiable
  Programming workshop at Neural Information Processing Systems 2021 (2021).

\bibitem{hughes2019forward}
T.~W. Hughes, I.~A. Williamson, M.~Minkov, S.~Fan, Forward-mode differentiation
  of {M}axwell’s equations, ACS Photonics 6~(11) (2019) 3010--3016.

\end{thebibliography}

\end{document}